**Toward a New Microscopic Framework for Kondo Lattice Materials**

Gilbert Lonzarich[1], David Pines[2,3], and Yi-feng Yang[4,5]

[1]Cavendish Laboratory, Department of Physics, Cambridge University, UK CB30HE
[2]Department of Physics, University of California, Davis, CA 95616
[3]Santa Fe Institute, Santa Fe, NM 87501
[4]Beijing National Laboratory for Condensed Matter Physics and Institute of Physics, Chinese Academy of Sciences, Beijing 100190, China
[5]Collaborative Innovation Center of Quantum Matter, Beijing 100190, China;

**Abstract**

Understanding the emergence and subsequent behavior of heavy electrons in Kondo lattice materials is one of the grand challenges in condensed matter physics. In this perspective we review the progress that has been made during the past decade and suggest some directions for future research. Our focus will be on developing a new microscopic framework that incorporates the basic concepts that emerge from a phenomenological description of the key experimental findings.

**I. Introduction**

Kondo lattice materials in which a lattice of localized spins are coupled to background conduction electrons are of great interest because they exhibit a range of remarkable collective electronic states that challenge our theoretical understanding of emergent quantum phenomena. However, the settings for such collective states can be enormously complex involving anisotropic orbital-dependent hybridization, multiplicity of nearly degenerate orbitals and intricate lattice structures that can play important roles in their low energy properties. To deal with this complexity there is a need for a viable phenomenological description together with a minimal microscopic model that can help to give meaning to the concepts underpinning the phenomenology.

In their discussion of emergent states and hybridization effectiveness in Kondo lattice materials, Yang and Pines have introduced a candidate for such a phenomenological description[1][2]. Their description is based on the concept of two interpenetrating fluids[3] that emerge below a coherence temperature in analogy to the two-fluid model of superfluidity (but in the absence of a true phase transition). In particular, their description is based on a small number of temperature scales; namely the coherence temperature, $T^*$, which represents a crossover from an incoherent high temperature state into a coherent two-fluid low temperature state; the delocalization temperature, $T_L$, which marks the crossover from the two-fluid state to a single fluid state; $T_{FL}$, which marks the crossover into a Fermi liquid state; and $T_N$, which marks a transition into a magnetically ordered state. In the vicinity of the quantum critical point where $T_N$ and $T_{FL}$ vanish as a function of a quantum tuning parameter, such as applied pressure, a new temperature scale can emerge, $T_c$, marking the emergence of an unconventional superconducting state (or other kind of exotic state).

A candidate for a minimal microscopic model that can help to clarify the nature of the two fluids and the temperature scales of the two-fluid description involves a single conduction band coupled to a lattice of singly occupied doubly degenerate f levels. This simplest Kondo lattice model ignores in particular the multiplicity of f levels and conduction electron levels that play an important role in many materials of practical interest. Nevertheless, it seems capable of illustrating the chief concepts behind the two-fluid model and we use it here.

For simplicity we focus mainly on quantum liquid states of the Kondo lattice that involve no long-range order, *e.g.*, states above any magnetic or superconducting phase transitions that are found in many heavy fermion materials at sufficiently low temperatures.

We begin by reviewing the conventional description of the Kondo interaction between localized f-electron spins and the background of conduction electron spins, which leads to hybridization within a slave boson mean field theory that excludes intersite magnetic correlations. These correlations are then taken into account within a similar mean field theory by adding to the Kondo on-site spin interaction an effective intersite Heisenberg interaction between the localized spins. We will find that both of these models underestimate the degree of hybridization needed to account for the experimental data in a number of Kondo lattice systems and suggest that what is missing is a phenomenon we will refer to as collective hybridization.

We next review what experiment has taught us about the 'hydrogen atoms" of heavy electron physics, $CeRhIn_5$ and $CeCoIn_5$ for example, and the way in which the two-fluid description enables one to disentangle the behavior of the emergent itinerant heavy electrons and local moments below the coherence temperature, $T^*$, and show how hybridization changes the emergence of antiferromagnetic and superconducting order.

In Sec. IV, we consider further possible scenarios for the development of an improved microscopic theory, while in Sec. V we give a brief summary of what we have learned, before presenting our concluding remarks. In two Appendices we suggest experiments that can clarify the basic concepts that may need to be incorporated in future microscopic models and consider the possibility that what we learn about heavy electron emergence may also be applicable to the coupling between d-electron localized spins and p-band holes in the cuprate superconductors.

## II. Local Kondo and Intersite Heisenberg Interactions

In the simplest case the coupling between conduction electrons and f electrons is described by an on-site spin interaction of the form $Js_i \cdot S_i$, where $i$ is the site index, $J$ is the local Kondo interaction parameter and $s/2$ and $S$ correspond to the spin ½ operators for conduction and f electrons, respectively. It is assumed that a strong Coulomb interaction between two electrons in an f orbital on a given site essentially forbids double occupancy of that orbital in a low frequency description.

In the simple Kondo hybridization model, below a crossover temperature known as the Kondo temperature, $T_K$, the behavior of the conduction and f electrons becomes strongly correlated or coherent. In particular, the spins of conduction and f electrons on the same site tend to align themselves in spin-singlet states. This is described in terms of the build-up of the effective order parameter of the form $\langle f_s^+ c_{-s}^+ \rangle_i$ or $\langle f_s^+ c_s \rangle_i$, where $f_s$ and $c_s$ are the fermion field operators of spin $s$ (↑ or ↓) for f electrons and conduction electrons, respectively, on the same site $i$.

Well above $T_K$ the conduction electrons and f electrons are essentially decoupled in the sense that the f electrons behave as a collection of disordered local moments with entropy ln 2 per site ($k_B = 1$ everywhere). On the other hand, well below $T_K$ the developing order parameter defined above gives rise essentially to a hybridization of the f-electron and conduction electron states that may be described in terms of lower and upper sub-bands separated by an energy gap known as the Kondo hybridization gap. The Fermi volume in this state encloses ($n_f + n_c$) electrons per site, where $n_f$ and $n_c$ are, respectively, the average number of f electrons and conduction electrons per site (we assume $n_c \leq 1$).

In the simplest slave boson mean field description[4] the minimum indirect hybridization gap in the ground state is $2T_K$ and the direct gap is $2\sqrt{T_K/\rho}$ in the symmetric case of a flat conduction electron density of states, $\rho = 1/D$, for both spins defined in the energy range $-D$ to $D$. In this mean field model the Kondo temperature $T_K$ is of order $\rho^{-1}\exp(-1/\rho J)$ when $\rho J \ll 1$, and of order $J$ in the strong coupling limit $\rho J > 1$. This is the same as the Kondo temperature for the single Kondo impurity problem and represents the coherence temperature $T^*$ in this mean field model.

As indicated in Table I, however, this prediction is inconsistent with observation. In practice $T^*$ is often found to be much larger than the single ion Kondo temperature $T_K$ near to magnetic quantum critical points. Moreover the above mean field model, when extended to include not only the hybridization order parameter but also a magnetic order parameter, yields a critical value of the dimensionless coupling $\rho J$ separating the magnetic and non-magnetic ground states at the magnetic quantum critical point, which is consistently higher than observed (Table I and Sec. IV).

The slave boson mean field description includes only local on-site spin correlations and is said to be applicable strictly only in the limit of large orbital degeneracy, $N$. For finite $N$ (equal to two in the case considered here) intersite spin correlations emerge as corrections to the mean field model. These correlations might be incorporated approximately in a slave boson mean field theory by adding an additional term in the Kondo lattice Hamiltonian of the form $J_H S_i \cdot S_j$, where $i$ and $j$ are nearest-neighbour states and $J_H$ is a Heisenberg interaction parameter. In the simplest case $J_H$ is induced by the Kondo interaction itself and is of the Ruderman-Kittel-Kasuya-Yoshida (RKKY) form, *i.e.*, $J_H$ is of the order of $\rho J^2$. The inclusion of both Kondo and Heisenberg interactions over-describes the spin correlations in an exact solution, but might be instructive in a mean field treatment for low frequency processes. Also, the Kondo-Heisenberg model can deal with more general cases where $J_H$ is not governed by the Kondo interaction $J$ alone.

| Compound | $T^*$ (K) | $T_K$ (K) | $\gamma$ (mJ mol$^{-1}$K$^2$) | $J\rho$ | $J$ (meV) | $c$ |
|---|---|---|---|---|---|---|
| CeRhIn$_5$ | 20 ± 5 | 0.15 | 5.7 | 0.10 | 40 | 0.45 |
| CeCu$_6$ | 35 ± 5 | 3.5 | 8 | 0.15 | 43 | 0.49 |
| CeCu$_2$Si$_2$ | 75 ± 20 | 10 | 4 | 0.15 | 90 | 0.47 |
| CePb$_3$ | 20 ± 5 | 3 | 13 | 0.15 | 28 | 0.41 |
| CeCoIn$_5$ | 50 ± 10 | 6.6 | 7.6 | 0.16 | 49 | 0.55 |
| CePd$_2$Si$_2$ | 40 ± 10 | 9 | 7.8 | 0.17 | 51 | 0.41 |
| CePd$_2$Al$_3$ | 35 ± 10 | 10 | 9.7 | 0.18 | 43 | 0.40 |
| CeRu$_2$Si$_2$ | 60 ± 10 | 20 | 6.68 | 0.19 | 66 | 0.42 |
| U$_2$Zn$_{17}$ | 20 ± 5 | 2.7 | 12.3 | 0.15 | 29 | 0.41 |
| URu$_2$Si$_2$ | 55 ± 5 | 12 | 6.5 | 0.17 | 62 | 0.45 |
| UBe$_{13}$ | 55 ± 5 | 20 | 8 | 0.19 | 57 | 0.43 |
| UPd$_2$Al$_3$ | 60 ± 10 | 25 | 9.7 | 0.21 | 51 | 0.48 |
| YbRh$_2$Si$_2$ | 70 ± 20 | 20 | 7.8 | 0.19 | 58 | 0.53 |
| YbNi$_2$B$_2$C | 50 ± 5 | 20 | 11 | 0.21 | 44 | 0.47 |

Table I. Experimental values of $T^*$, the single ion Kondo temperature, $T_K$, and linear coefficient of the specific heat for the conduction electrons, $\gamma$, for a variety of Kondo lattice compounds. Also given are (i) values of $\rho J$ and $J$ inferred from $T_K$ and $\gamma$, and (ii) values of the ratio $c = T^*/\rho J^2$. Our definitions of $\rho$ and $J$ are given in Sec. II. For the method of determination of $T_K$ and $\gamma$ and for the original references see Yang et al. 2008.[1] Note that most of these materials are near quantum criticality, as their $\rho J$ values are close to the quantum critical value [see text] of $\rho J_c = 0.17$.

This more general Kondo-Heisenberg framework allows us to consider a very different limiting description of the normal state in which the Heisenberg rather than the Kondo interaction might appear to play the dominant role in a mean field description. In this limiting case the crossover from the disordered local f-electron state at high temperature and the coherent state at low temperature is initially governed by the build-up of intersite correlations between f electrons rather than hybridization of f electrons with conduction electrons.

The intersite f-electron interactions tend to promote nearest neighbor f-spin singlets below a coherence temperature $T^*$, which can be of the order of $J_H$ rather than $T_K$ in the limit being considered. The hybridization of conduction electrons and f electrons in this case begins at still lower temperatures and becomes essentially well developed below a scale we might naively associate with the delocalization temperature in the two-fluid model, $T_L < T^*$.

Thus in this picture the intersite singlet order parameter of the form $\langle f_{is}^+ f_{j,-s}^+ \rangle$ or $\langle f_{is}^+ f_{js} \rangle$ develops below $T^*$, while the on-site singlet order parameter $\langle f_{is}^+ c_{i,-s}^+ \rangle$ or $\langle f_{is}^+ c_{is} \rangle$ tends to develop at a still lower temperature, $T_L$.

In a slave boson mean field model similar to that described above, but now including both the Kondo and Heisenberg interactions, the order parameter $\langle f_{is}^+ f_{js} \rangle$ leads to an effective dispersion of the f band (the spinon band) for $T \leq T^*$, while the order parameter $\langle f_{is}^+ c_{is} \rangle$ leads to a hybridization of the f band and conduction band for $T \leq T_L < T^*$.[4] In this limit the coherence temperature is of the order of $J_H$.

In this mean field model in which the Heisenberg interaction plays a dominant role the coherence temperature at a magnetic quantum critical point can in principle greatly exceed the single ion Kondo temperature, as observed experimentally (Table I). However, this model exhibits a hybridization gap only below $T_L$ and hence well below $T^*$, a prediction inconsistent with the observation of the formation of a pseudogap in the optical conductivity at temperatures as high as $T^*$ and beyond (Sec. III.3). Within the Kondo-Heisenberg model a pseudogap at temperatures of order $J_H$ might be expected to arise in principle even without hybridization from the formation of intersite spin singlets described by an order parameter of the form $\langle f_{is}^+, f_{i,-s}^+ \rangle$, or due to short range magnetic order in a description beyond the mean field approximation.

Another possibility is that the pseudogap above $T_L$ arises from an enhanced form of hybridization, which is missed in the slave boson mean field approximation. We argue in Sec. IV via a different type of mean field theory that hybridization in a lattice might indeed be greatly enhanced compared with that for a single ion. This lattice-enhanced hybridization will be referred to as "collective hybridization".

In such a revised description including collective hybridization we imagine that the Kondo liquid state in the temperature range between $T_L$ and $T^*$ might be characterized by slow spatial fluctuations in the hybridization field leading to a mosaic of momentarily hybridized and unhybridized states. The hybridized regions can be described in terms of itinerant fermion quasiparticles composed of a coherent superposition of f electrons and conduction electron states, and the unhybridized regions in terms of local f-electron moments along with a separate system of conduction electrons weakly coupled with the local moments. This leads to a picture in which, in some sense, itinerant heavy quasiparticles coexist with essentially local moments, potentially forming the two "fluids" in the two-fluid model for the intermediate temperature range between $T_L$ and $T^*$.

**III. Experiments that measure the emergent physical behavior of CeRhIn$_5$ and CeCoIn$_5$ and a phenomenological description that explains these.**

**1. The phase diagrams**

Changes in temperature, pressure and external magnetic fields bring about the changes in the physical behavior of the prototypic and most-studied Kondo lattice materials, CeRhIn$_5$ and CeCoIn$_5$, that are pictured in the schematic phase diagram given in Fig. 1 and the experimental phase diagrams shown in Fig. 2. In Fig. 1 we see that at high temperatures we have two independent components: a Kondo lattice of f-electron local moments that form a spin liquid, and a background of conduction electrons. Matters change below a crossover temperature, $T^*$, usually called the coherence temperature, as itinerant heavy fermions begin to emerge as a result of the hybridization of the localized f-electron moments with the background conduction electrons. Below $T^*$ we find two new components whose physical properties can be disentangled in the ways discussed in the following sections: one component is the local moment spin liquid whose properties are to some degree affected by hybridization; the second is an unusual Fermi liquid formed by the emergent itinerant heavy electrons. (Unhybridized conduction electrons are also present, but these play little role in determining the physical behavior of the material.)

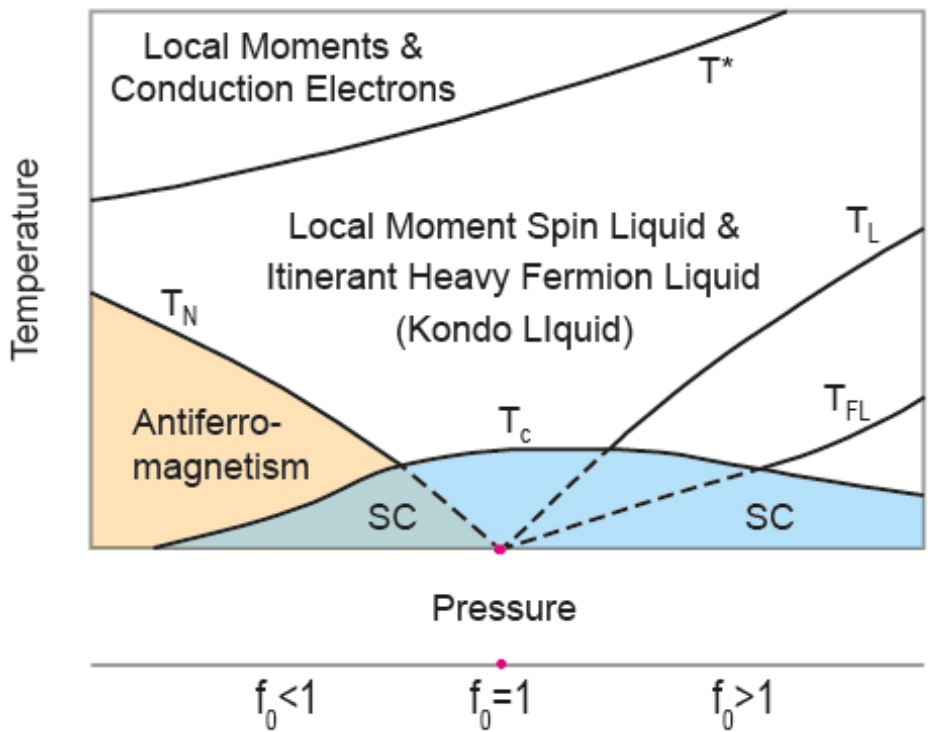

Fig. 1. A schematic phase diagram for heavy electron superconductors that is based on pressure-induced changes in the behavior of $CeCoIn_5$ and $CeRhIn_5$.[5][6] It shows the coherence temperature, $T^*$ ($> 30 T_c$), at which itinerant heavy electrons emerge to form a Kondo liquid, the delocalization line, $T_L$, below which local moments are suppressed and one finds a different anomalous "quantum critical" Fermi liquid, the $T_{FL}$ line that marks the crossover to a Landau Fermi liquid behavior, and the low temperature ordered phases of the local moments and the Kondo liquid, as a function of a quantum tuning parameter, $f_0$, that provides a quantitative measure of hybridization effectiveness in the two-fluid model. $T_N$ is the Néel temperature and $T_c$ is the superconducting (SC) transition temperature. Shown is a special case in which the $T_N$ and $T_L$ lines meet at a quantum critical point (solid red point). In general, the quantum critical points at which magnetism and delocalization begin to develop occur at different pressures.

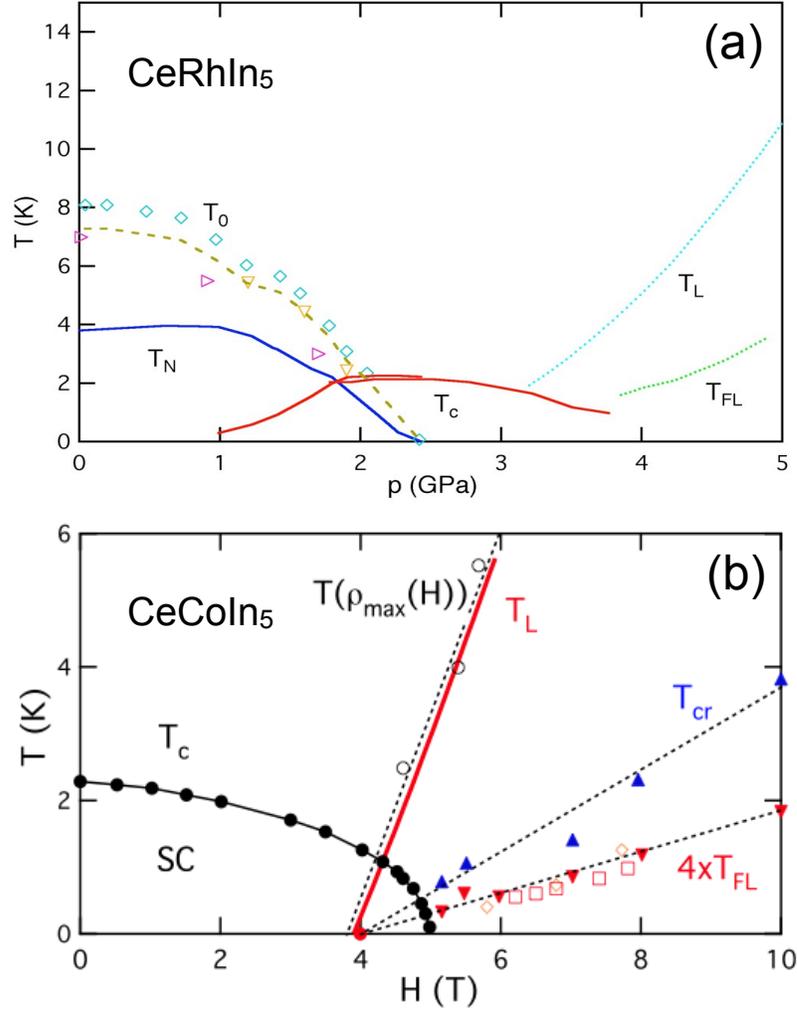

Fig. 2. A sketch of the normal state phases observed near quantum critical points; (a) near the magnetic quantum critical pressure for CeRhIn$_5$ [12]; (b) near the magnetic quantum critical magnetic field of CeCoIn$_5$.[10]. To the left of the delocalization line, $T_L$, one has both a Kondo liquid and local moment spin liquid present; to its right, no local moments are present, and in place of the Kondo liquid with its universal scaling behavior based on $T^*$, three regimes of the heavy electron liquid have been identified separated by the crossover lines $T_L$, $T_{FL}$, already defined, and an additional scale $T_{cr}$ as discussed in references [13][14][15].

A number of experiments have shown that the emergent heavy electron liquid does not behave as a standard Landau Fermi liquid; it is instead a new state of matter with a bulk specific heat and spin susceptibility that display universal scaling behavior with a scale that is set by $T^*$ and a density of quasiparticle states that near the quantum critical point is proportional to $\ln(T^*/T)$. In view of its unique universal properties, and as a way of distinguishing it from Fermi liquid phases that display different behavior, it has frequently been called a Kondo liquid, a notation we adopt here.

Both the local moment spin liquid and the Kondo liquid order at low temperatures; for a wide range of pressures the local moments develop long-range antiferromagnetic order, while the co-existing Kondo liquid becomes an unconventional superconductor. Moreover, as the pressure increases superconductivity becomes increasingly robust; when $T_c > T_N$, local moment order becomes suppressed, and one finds only the superconducting state.

The experiments on which Fig. 1 is based show that there is a pressure and temperature below which the hybridization process is essentially complete and the local moment spin liquid is suppressed. This defines a pressure-dependent delocalization line, $T_L$, that terminates at a quantum critical point at $T = 0$. This quantum critical point can coincide with the magnetic quantum critical point defined by $T_N = 0$, as shown in Fig. 1, but may also lie within the antiferromagnetic (AF) dome or outside it. The maximum in the superconducting transition temperature is found to occur at, or close to, the intersection of $T_L$ and $T_c$ in the materials discussed here.

Fig. 1 shows the principal transition or crossover lines relevant to our discussions. Additional crossover lines have also been identified and shown, for example, in Fig. 2. Fig. 2a includes a crossover line labelled $T_0 > T_N$ and tracking the $T_N$ line. Below $T_0$ collective hybridization appears to diminish in a relocalization process discussed elsewhere.[7]

Pressure is not the only way to alter the physical behavior of Kondo lattice materials and bring about quantum criticality; this can also be accomplished by doping[8] or by applying strong magnetic fields[10]. In Fig. 2b we give a sketch of the behavior that is seen experimentally[11] for $CeCoIn_5$ in the vicinity of a magnetic quantum critical point that lies just within the superconducting dome.

Besides $T_L$ and $T_{FL}$ an additional crossover temperature, $T_{cr}$, has been identified by Zaum et al.[11], which is said to separate regimes with different non-Fermi liquid characteristics. These regimes might in principle emerge below $T_L$ in the absence of a magnetic field, but these have not yet been identified, either because these require large magnetic fields in order to emerge or because their presence may be masked by superconductivity.

The physical origin of the normal state phase changes shown in Fig. 1 is the increase in the effectiveness of hybridization as the pressure is increased, as is clear from the positive slope of $T_L$. Fig. 2b shows that application of strong magnetic fields also leads to an increase in hybridization effectiveness for $CeCoIn_5$. One of the virtues of the phenomenological two-fluid model we now consider is that it introduces a hybridization effectiveness parameter that enables one to quantify both pressure and magnetic field induced changes in hybridization.

## 2. Hybridization and a phenomenological description of the two coexisting fluids in the normal state of heavy electron materials

*T\* and the collective origin of hybridization.* The issue of the physical origin of the hybridization between local moments and conduction electrons that is responsible for heavy electron emergence was resolved in a 2008 review of the experimental literature by Yang, Fisk et al.[1], hereafter YF. If *T\** reflected single-ion hybridization, as in the Kondo effect for isolated magnetic impurities, it would be given by the single-ion Kondo temperature

$$T_K = \rho^{-1} \exp(-1/\rho J) \qquad [1]$$

where the local Kondo coupling $J$ and the bare conduction electron density of states $\rho$ are as defined in Sec. II. In a study of some 14 Kondo lattice materials for which $T_K$ can be determined experimentally by diluting the concentration of localized f electrons, YF found that $T^*$ is always large compared to $T_K$, and that it is governed instead by the nearest neighbor local moment intersite coupling $J_H$ and more precisely the RKKY interaction[16][17][18]. We can therefore write

$$T^* = cJ^2\rho \qquad [2]$$

where $c$ is a constant determined by the details of the hybridization and the conduction electron Fermi surface. In Fig. 3 we reproduce the results of the analysis of $T^*$ and $T_K$ that led YF to conclude that $c \sim 0.45$ for the remarkably wide range of heavy electron materials shown there, and that the local moment hybridization is a *collective* phenomenon in the sense defined in Sec. II and Sec. IV.

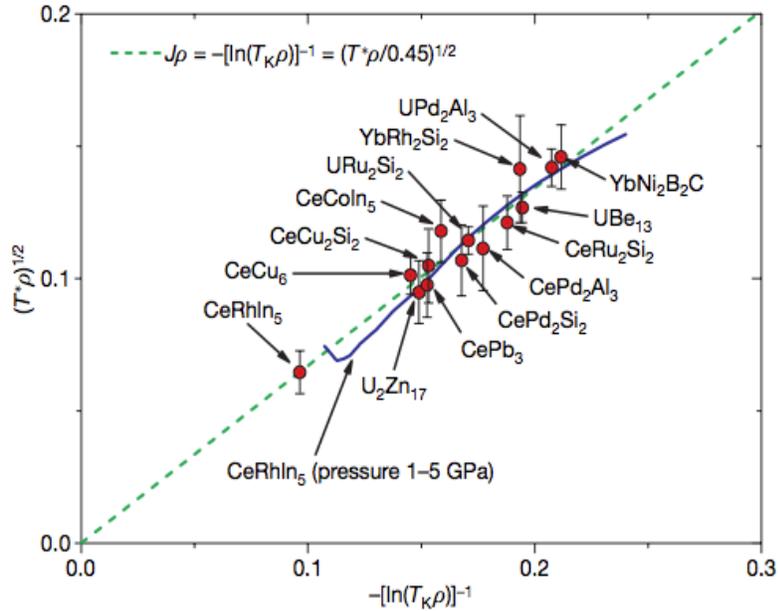

Fig. 3. The experimental confirmation for a variety of Kondo lattice materials that $T^*$ is given by the intersite RKKY interaction. The solid line shows $T^*$ (resistivity peak) of CeRhIn$_5$ under pressure from 1 GPa (lower left) to 5 GPa (upper right). Each error bar indicates a typical range of uncertainties for estimating $T^*$. The dashed line is a guide to all the materials. The evident correlation between $T^*$, $T_K$ and $\rho$ provides verification of the RKKY nature of $T^* = c\rho J^2$. Small deviations from the $c \sim 0.45$ line may result from details of the hybridization and the conduction electron Fermi surface.[1]

*A two-fluid description.* Although heavy electrons in f-electron systems were discovered in the 1970s, it is only within the past decade that it has proved possible to disentangle the effects of interactions between conduction and f electrons and to identify universal scaling behavior of what we have called the Kondo liquid state. The first step in this direction was taken by Nakatsuji *et al.*[3], who introduced a phenomenological two-fluid description to explain the results of bulk magnetic susceptibility and specific heat experiments carried out on $Ce_{1-x}La_xCoIn_5$, a material that is particularly well suited for investigation because its crystal electric field splitting is at an energy well above the coherence temperature, $T^*$, at which heavy electrons first emerge, so plays no complicating role in the heavy electron behavior. By analogy with superfluid $He_4$, they introduced a phenomenological "order parameter", $f(T)$, to describe the strength of the emerging itinerant heavy fermion liquid, and proposed that below the coherence temperature, $T^*$, the bulk susceptibility would take a two-fluid form () that we would now write as:

$$\chi = f(T/T^*)\chi_{KL} + \left[1 - f(T/T^*)\right]\chi_{SL} \qquad [3]$$

where $\chi_{KL}$ is the intrinsic susceptibility of the itinerant heavy fermion liquid, the Kondo liquid (KL), and $\chi_{SL}$ is that of the local moment spin liquid (SL) defined in Sec. III 1, with a similar expression for the bulk specific heat, $C$. They found that a reasonable fit to their experimental data could be obtained by taking the Wilson ratio between $\chi_{KL}$ and $C_{KL}/T$ to be of order two and the phenomenological order parameter to be of the form $f(T)=(1-T/T^*)$. They further noted that $T^* \sim 45$ K was of the order of the intersite coupling and large compared to the single ion Kondo temperature, 1.4 K.

Soon afterwards, their work was confirmed and expanded by Curro *et al.* [CYSP][19] in a detailed analysis of Knight shift experiments on a wide variety of heavy electron materials. CYSP noted that the so-called hyperfine anomaly, in which the Knight shift no longer tracked the bulk susceptibility, could be understood as marking the emergence of the itinerant heavy fermion component to the susceptibility. They showed how this could be obtained directly from experiment and, for some fourteen heavy electron materials, found that it took the form proposed by Nakatsuji *et al*[3].

*The Kondo liquid.* Additional experimental information and a careful re-analysis of the magnetic data then led Yang and Pines [YP][20] to conclude that the emergent itinerant heavy electron liquid would exhibit universal behavior in the two-fluid regime between $T^*$ and an ordering temperature or $T_L$ near to the quantum critical point as in Fig. 1. They proposed that it represented a new state of quantum matter, which they called a Kondo liquid. They found that a better fit to experiment could be obtained with (i) an order parameter that differed slightly from that proposed by NPF,

$$f(T/T^*) = f_0(1 - T/T^*)^{3/2} \qquad [5]$$

where $f_0$ is the hybridization effectiveness discussed below and (ii) a quasiparticle density of states at the Fermi level for the Kondo liquid proportional to

$$\rho_{KL} = \left(1 - T/T^*\right)^{3/2}\left[1 + \ln\left(T/T^*\right)\right] \qquad [6]$$

We note that this determines the contribution of the Kondo liquid component to the total entropy if we take the local moment spin liquid component to be ln2 per moment. A review of the two-fluid model is given by Yi-feng Yang in this issue.

As may be seen in Fig. 4, this expression provided an excellent fit to the values of the anomalous Knight shift obtained from NMR or μSR experiments for the 115 family of heavy electron materials and all others that display a Knight shift anomaly. They then showed that, three other experiments on the 115 materials provided additional evidence for the emergence at *T\** of a new state of matter whose density of states was given by Eq. (6): the anomalous Hall effect measured by Hundley *et al.*[21], a conduction asymmetry in tunneling experiments discovered by Park *et al.*[22], and the Fano asymmetry parameter measured in Raman scattering experiments by Martinho *et al.*[23].

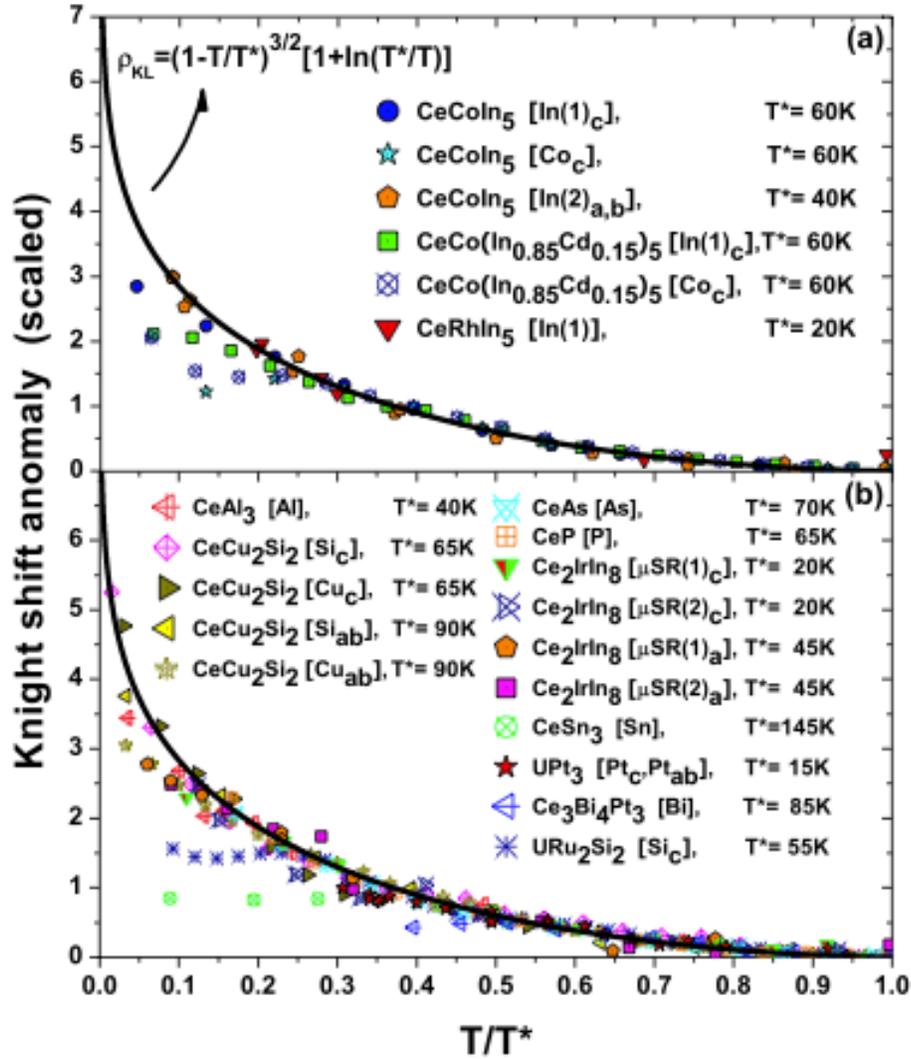

Fig. 4. Scaling behavior of the anomalous Knight shift for Ce115 and some other heavy-electron materials.[20]

*A connection to microscopic theory.* Remarkably, as discussed by YP, a connection could be established between their phenomenological description and microscopic theory through the seminal microscopic calculations of Shim *et al.* [SHK][24]. SHK carried out a first principles calculation of the localized to itinerant transition in CeIrIn$_5$ using Dynamical Mean Field Theory in combination with Local Density Approximation (LDA+DMFT) that implicitly allows for feedback and collective effects. In Fig. 5 their calculated density of states at the quasiparticle peak for CeIrIn$_5$ is compared with the YP universal KL density of states. The excellent agreement between the two results indicates that SHK have captured the Kondo liquid state density, a truly remarkable finding that provides quite strong evidence for the validity of both their microscopic approach to hybridization and the YP phenomenological scaling expression for the KL state density.

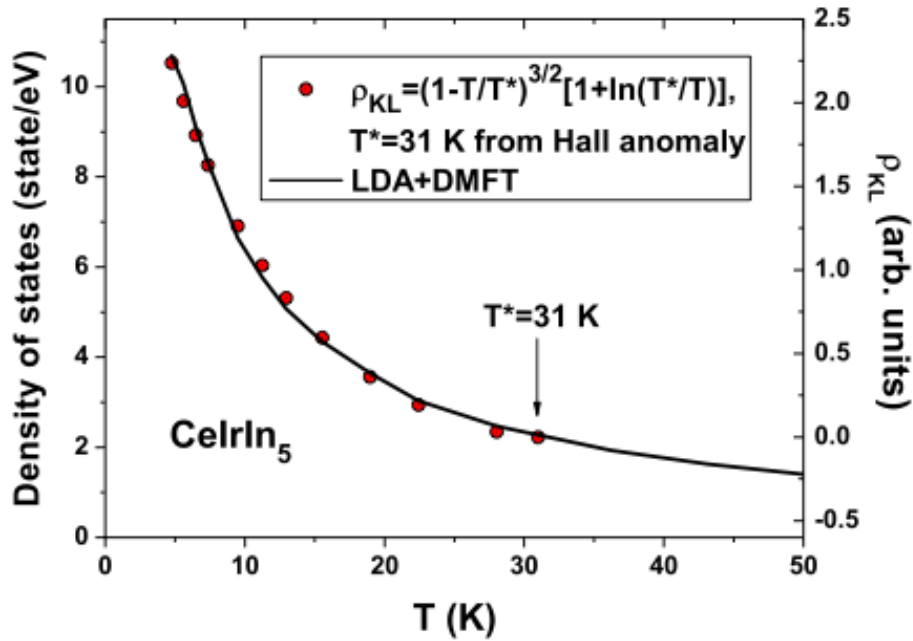

Fig. 5. The calculated LDA+DMFT density of states of the quasiparticles in CeIrIn$_5$[24] compared to the YP universal Kondo liquid (KL) density of states. The coherence temperature, $T^* = 31$ K is fixed by experiments on the emergence of the Hall anomaly.[20]

*Hybridization effectiveness.* Yang and Pines [YP1][2] then showed how a careful study of the role played by the pressure dependent hybridization effectiveness parameter, $f_0$, that appears as a prefactor in the heavy electron order parameter,

Eq. (5), enables one to obtain the conditions necessary for the emergence of various low temperature ordered states in Kondo lattice materials whose dependence on $f_0$ is sketched in Fig. 1. The normalization of $f_0$ is set by the requirement that it be unity at the quantum critical pressure at which the local moments have completely hybridized at $T = 0$; as YF note its magnitude elsewhere is best determined from measurements of the bulk magnetic susceptibility and find that it approximately scales with $T^*$ as the pressure is varied.

It follows that if $f_0 < 1$ one will have present local moments that order antiferromagnetically. On the other hand, if $f_0 > 1$, Eq. (5) tells us that one reaches $f(T) = 1$ along a pressure-dependent temperature line, $T_L(P)$,

$$T_L = T^* \left(1 - f_0^{-2/3}\right)$$

[7]

below which local moment behavior is suppressed and one has a single component for which Eq. (6) no longer applies. At sufficiently low temperatures, below $T_{FL}$, the itinerant heavy fermions may be described in term of the Landau Fermi liquid theory. $T_L$ (along with $T^*$ and $T_{FL}$) are crossover temperatures and in a more realistic model $f(T)$ would be taken to be a smooth function of $T$.

### 3. The hybridization gap

So far we have focused on magnetic experiments that tell us about the emergence and subsequent behavior of the Kondo liquid in the spin channel. Important additional information about its emergence and subsequent behavior is provided by optical experiments that probe its behavior in the charge channel. These reveal a rapid development of a hybridization gap below $T^*$. As shown in Fig. 6 for $CeCoIn_5$, part of the spectral weight of the broad Drude peak of the conduction electrons is pushed away at intermediate frequencies, leading to a narrow Drude peak at low frequencies and a broad hump at the mid-infrared frequencies (mid-IR peak). These correspond to the opening of an indirect gap and a direct gap in the renormalized band structures due to hybridization, as illustrated in Fig. 6.

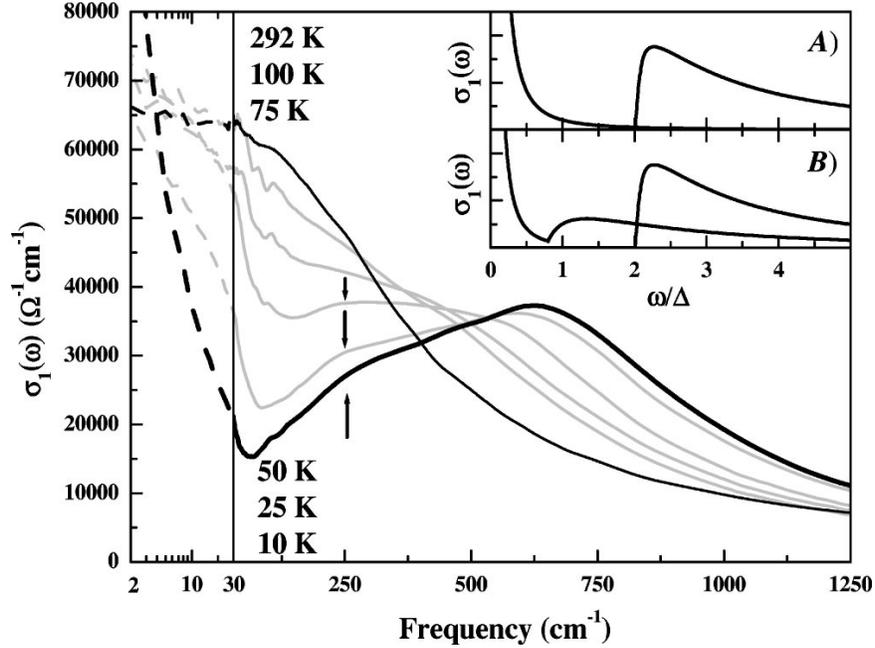

Fig. 6. The optical conductivity of $CeCoIn_5$. Evidence for the development of a hybridization gap with decreasing temperature.[25]

As discussed in Sec. II and more fully in Sec IV, mean-field theory calculations predict that the direct gap should be related to the square root of the single ion Kondo temperature $\Delta \sim \sqrt{T_K/\rho}$. However, in a number of materials in which $T_K$ has been measured, this expression seems to underestimate the direct gap.

On the other hand, if one replaces the single ion Kondo temperature $T_K$ with the energy scale that characterizes the Kondo liquid, $T^*$, we find that

$$\Delta \sim \sqrt{T^*/\rho} \sim J \qquad [8]$$

yields agreement with experiment for a number of heavy electron materials as may be seen in Fig. 7.

Fig. 6 also shows that signatures of the hybridization gap are already seen above $T^*$. This suggests that the onset of the collective hybridization that yields the Kondo liquid is a crossover rather than a well-defined phase transition.

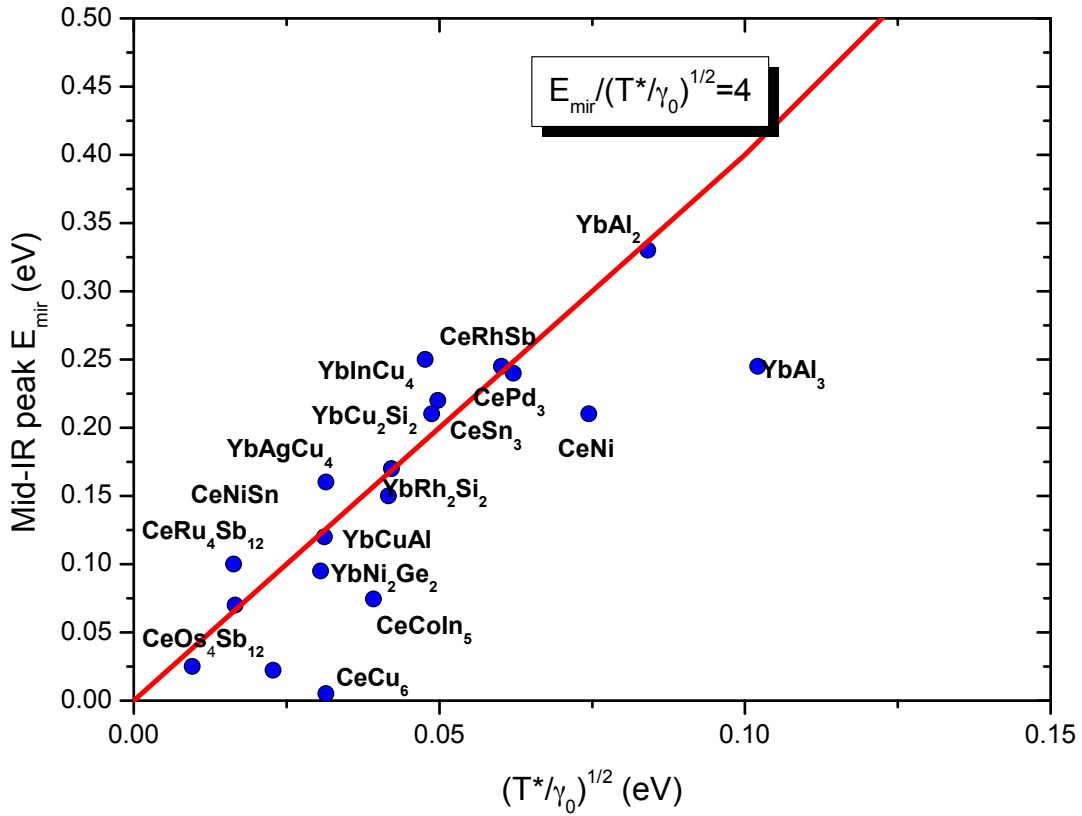

Fig. 7. Comparison of the mid-IR peak energy with $\sqrt{T^*/\rho}$. The conduction electron density of states at the Fermi level, $\rho$, is determined from the linear coefficient of the heat capacity, $\gamma_0$, expressed in units of inverse bandwidth, of a close relative compound with no f electrons.[1][26]

## 4. Fermi surface investigations

The above studies have greatly clarified the nature of the coherent state of Kondo lattice systems and have emphasized the importance of itinerant heavy fermions arising from the effect of hybridization between conduction electrons and f electrons. The existence of well defined elementary excitations in the form of itinerant heavy fermions and their detailed properties at low temperatures have been established in particular by means of measurements of quantum oscillations of the magnetization and electrical resistivity in numerous Kondo lattice systems.

An early example is the heavy fermion paramagnet $CeRu_2Si_2$ and its ferromagnetic isoelectronic and isostructural partner $CeRu_2Ge_2$, which differs structurally from $CeRu_2Si_2$ only in having slightly increased lattice spacing.[27][28] Quantum oscillation studies show that the Fermi surface in the ferromagnetic state of $CeRu_2Ge_2$ corresponds to that expected for the conduction electrons alone, but with a small exchange splitting and mass renormalization attributed to a weak coupling to the localized f electrons. On the other hand, the Fermi surface of the paramagnetic state of $CeRu_2Si_2$ corresponds to that expected for the case in which not only the conduction electrons, but also the f electrons (hybridized with the conduction electrons) are itinerant. The quasiparticle excitations across the large Fermi surface are coherent superpositions of conduction and f-electron states with high effective masses that account for the observed linear heat capacity. Similar conclusions have been reached in more recent studies in a number of heavy electron systems and, in particular, the Ce115 compounds.[29][30]

These results apply within the paramagnetic phase or deep within the magnetic state where the conduction and f electrons are either fully hybridized or fully unhybridized (*i.e.*, to the right and well to the left, respectively, of the quantum critical point in Fig. 1). There is also considerable experimental evidence that magnetic order and hybridization can coexist[30][31], in which case the $T_L$ line tends to zero within the magnetic state in the two-fluid model. This implies the existence of separate quantum critical points associated with magnetic order and hybridization (with $T_L$ and $T_N$ vanishing at different pressures in Fig. 1). This scenario is consistent with recent numerical analyses of the Kondo lattice model[32] and has been the basis of analytical descriptions of the low temperature properties of systems such as $YbRh_2Si_2$[33] on the border of magnetic quantum critical points.

The nature of the Fermi surface in the two-fluid regime above $T_N$ or $T_{FL}$ remains unclear. One possibility is that the Fermi surface is not well defined in this regime in the sense suggested in Sec. II and for example in ref. [34].

**5. The influence of hybridization on local moment antiferromagnetism and Kondo liquid superconductivity**

The two-fluid model enables one to quantify the effects of hybridization on local moment antiferromagnetism and Kondo liquid superconductivity by combining information on $T^*$ and $f_0$ from transport and magnetic experiments with toy models for the emergence of these ordered states.

As discussed in YP, the magnetic properties of the local moment spin liquid can be described by a mean-field expression for its dynamic susceptibility,

$$\chi(q,\omega) = \frac{f_l \chi_0}{1 - J(q) f_l \chi_0 - i\omega / \gamma_l}$$

[9]

where $f_l(T) = (1 - f(T))$ is a relative strength, $\chi_0$ is the local susceptibility of an individual f moment, $\gamma_l$ is the local relaxation rate, $J(q)$ is the Heisenberg or RKKY exchange coupling, and the magnetic moment of the f-spins have been set to unity. The local moment spin liquid will then begin to order at a Néel temperature defined by

$$J(q^*) f_l(T_N) \chi_0(T_N) = 1$$

[10]

where *q\** is the ordering wavevector. If we take for simplicity $\chi_0 = C/T$, where *C* is the Curie constant, then

$$\frac{T_N}{T^*} = \eta_N f_I(T_N)$$ [11]

where the parameter $\eta_N = CJ(q^*)/T^*$ reflects the effect of the RKKY interaction, while $f_I(T_N)$ reflects the role played by collective hybridization in reducing the Néel temperature. (This will be discussed more fully in Sec. IV2.)

Figs 8c and 8d shows that this simple mean field model yields excellent agreement with experiment for the variation with doping of the Néel temperature of Cd-doped CeCoIn$_5$ and its variation with pressure for CeRhIn$_5$.

Recently, two of us [hereafter YP3] [35] have proposed a phenomenological "toy" BCS-like model for spin-fluctuation induced superconductivity in the 115 materials. For phonon-induced superconductivity, the BCS expression for $T_c$ depends on three quantities: the average energy range, $T_m$, over which the phonon-induced interaction between quasiparticles is attractive, the quasiparticle density of states at the Fermi level, $N_F$, and the average strength, *V*, of that attraction. The YP3 phenomenological heavy electron quantum critical magnetic analogue involves only these same three quantities and takes the form:

$$T_c(p) = 0.14 T_m^* \exp\left(-\frac{1}{N_F(p,T_c)V(p)}\right) = 0.14 T_m^* \exp\left(-\frac{1}{\eta\kappa(p)}\right)$$ [12]

where the range of energies over which the quantum critical spin-fluctuation induced interaction will be attractive is assumed to be proportional to $T_{QC}^*$, the coherence temperature at the quantum critical pressure, the constant prefactor 0.14 is determined from experiment, $N_F(p,T_c)$ is the heavy electron density of states at $T_c$ that depends on the region in which superconductivity emerges, and $V(p)$ is the average strength of the effective heavy electron spin-fluctuation induced attractive quasiparticle interaction. (We note that $V(p)$ is a phenomenological quantity that can be imagined to formally include strong coupling corrections expected to be important in heavy electron systems.)

YP3 make the physically plausible assumption that $V(p)$ scales with the pressure dependent interaction between local moments, which scales as $T^*(p)$, and so takes the form, $V(p) = \eta T^*(p)$, where $\eta$ is a material-dependent parameter that measures the effectiveness of spin fluctuations in bringing about superconductivity. Since the Kondo liquid specific heat varies inversely as $T^*$, for a given choice of $\eta$, the pressure dependence of $T_c$ is determined essentially by the dimensionless pairing strength

$$\kappa(p) = T^*(p) N_F(p, T_c) \qquad [13]$$

whose values can be determined for a given material by applying the two-fluid model.

In both CeRhIn$_5$ and CeCoIn$_5$ superconductivity arises from the Kondo liquid under three quite different regimes of hybridization: (i) where the hybridization is weak, it emerges in the presence of ordered local moments; (ii) around the quantum critical pressure, it emerges in the presence of the local moment spin liquid; (iii) at high pressures, where no local moments are present, it emerges from the itinerant heavy fermion liquid alone. These different regimes give rise to the variation in the dimensionless pairing strength $\kappa(p)$ shown in Fig. 8a. Quite remarkably, and unexpectedly, as shown in the remainder of Fig. 8, this simple toy model that contains only one free pressure independent parameter, $\eta$, provides an excellent fit to the pressure dependence of $T_c$ in both materials.

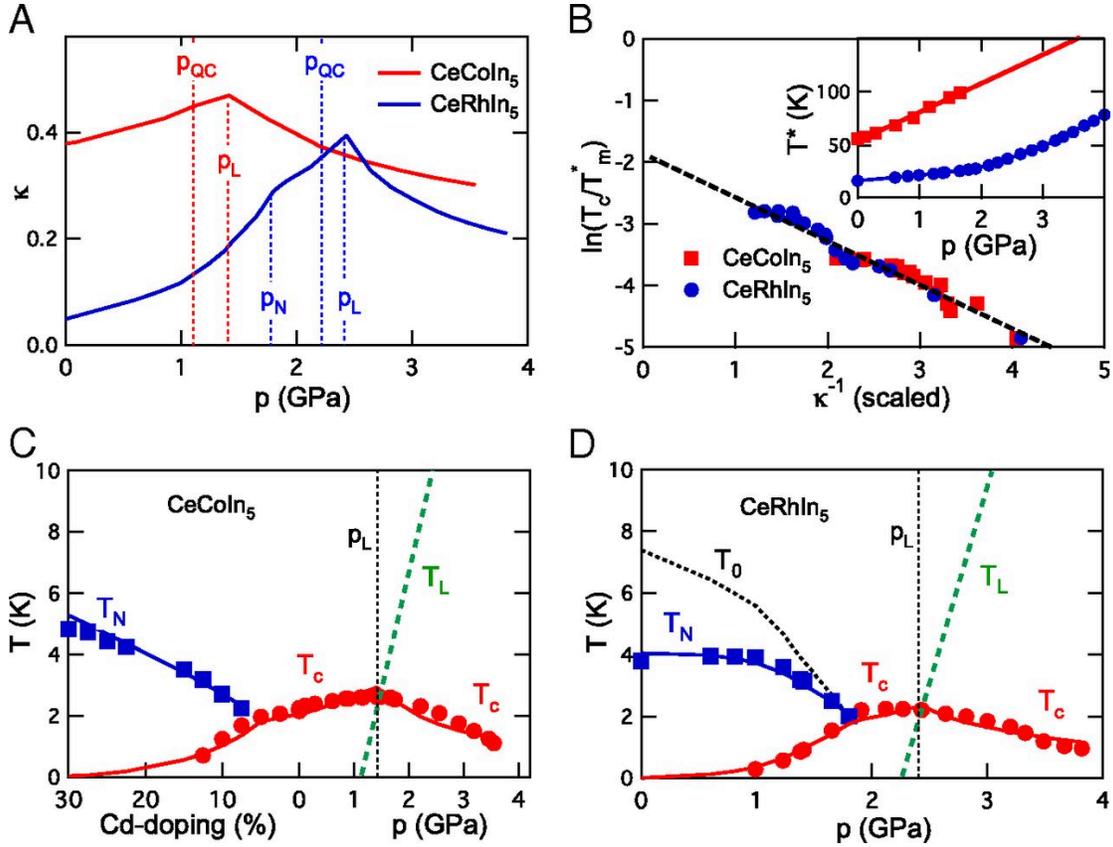

Fig. 8. Comparison of theory and experiment for the ordering temperatures measured in $CeCo(In_{1-x}Cd_x)_5$ and $CeRhIn_5$. (A) Pressure variation of the predicted dimensionless pairing strength, $\kappa(p)$. (B) Variation of $\ln(T_c/T_m^*)$ vs $\kappa(p)^{-1}$ (scaled) for $CeCoIn_5$ and $CeRhIn_5$. (Inset) The experimental values of $T^*(p)$ that are used to obtain $\kappa(p)$ in both compounds. (C) Comparison of the predicted (solid lines) and experimental $T_c$ and $T_N$ in $CeCo(In_{1-x}Cd_x)_5$ and $CeCoIn_5$. (D) Comparison of the predicted (solid lines) and experimental $T_c$ and $T_N$ in $CeRhIn_5$.[35]

**6. Changes in hybridization effectiveness brought about by quantum criticality**

We have seen in Fig. 2 that changes in pressure and magnetic field can bring materials such as $CeCoIn_5$ and $CeRhIn_5$ to the quantum critical point and have suggested that the physical origin of such changes might be pressure and field induced changes in hybridization effectiveness. Support for this view comes from a simple unified model for the combined influence of pressure and magnetic field on the two-fluid hybridization effectiveness parameter, $f_0$, in the vicinity of a quantum critical point. It leads to quantum critical and delocalization lines that accord well with those measured for CeCoIn$_5$ and yields a quantitative explanation of the field and pressure-induced changes in antiferromagnetic ordering and quantum critical behavior measured for YbRh$_2$Si$_2$. [hereafter YP4][36]

YP4 assume that

$$f_0(p,H) = f_0(p)\left[1+(\eta H)^\alpha\right] = \left[1+\eta_p\left(p-p_c^0\right)\right]\left[1+(\eta H)^\alpha\right] \quad [14]$$

where the exponent, $\alpha$, allows for magnetic field induced scaling behavior. With suitable choices of $\eta_p$ and $\eta$, they find that using the two-fluid model, with $\alpha = 2$ for CeCoIn$_5$ and $\alpha = 0.8$ for YbRh$_2$Si$_2$, a good account can be given of:

* The $T_L$ and $T_N$ lines and the pressure vs. magnetic field quantum critical line, as seen in Figs 9 and 10;

* The magneto-resistivity and spin-lattice relaxation rates of CeCoIn$_5$ at pressures near the quantum critical pressure and

* The field dependence of the specific heat, Knight shift, spin-lattice relaxation rate, and magnetoresistivity of YbRh$_2$Si$_2$ near its quantum critical magnetic field.

Their success in developing a phenomenological account of quantum critical scaling behavior suggests that for these phenomena the leading impact of quantum criticality is through its influence on hybridization effectiveness.

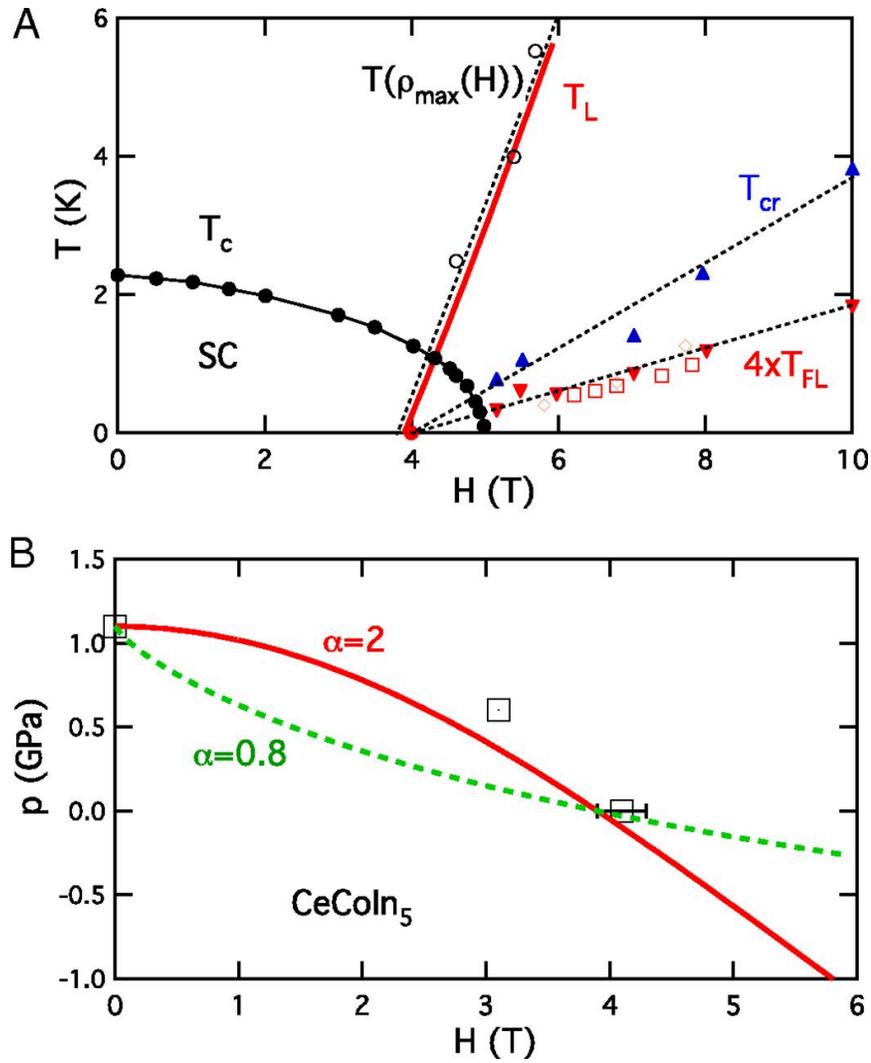

Fig. 9. Phase diagrams of CeCoIn$_5$. (A) Transition and crossover lines in the temperature-magnetic field plane. The symbols are as defined in the text and ref. [35]. (B) The quantum critical line (points of vanishing $T_{FL}$) in the pressure-magnetic field plane. Comparison of measurements (hollow squares) and the model discussed in the text (solid and dashed lines for different values of the phenomenological parameter $\alpha$) are shown.[35]

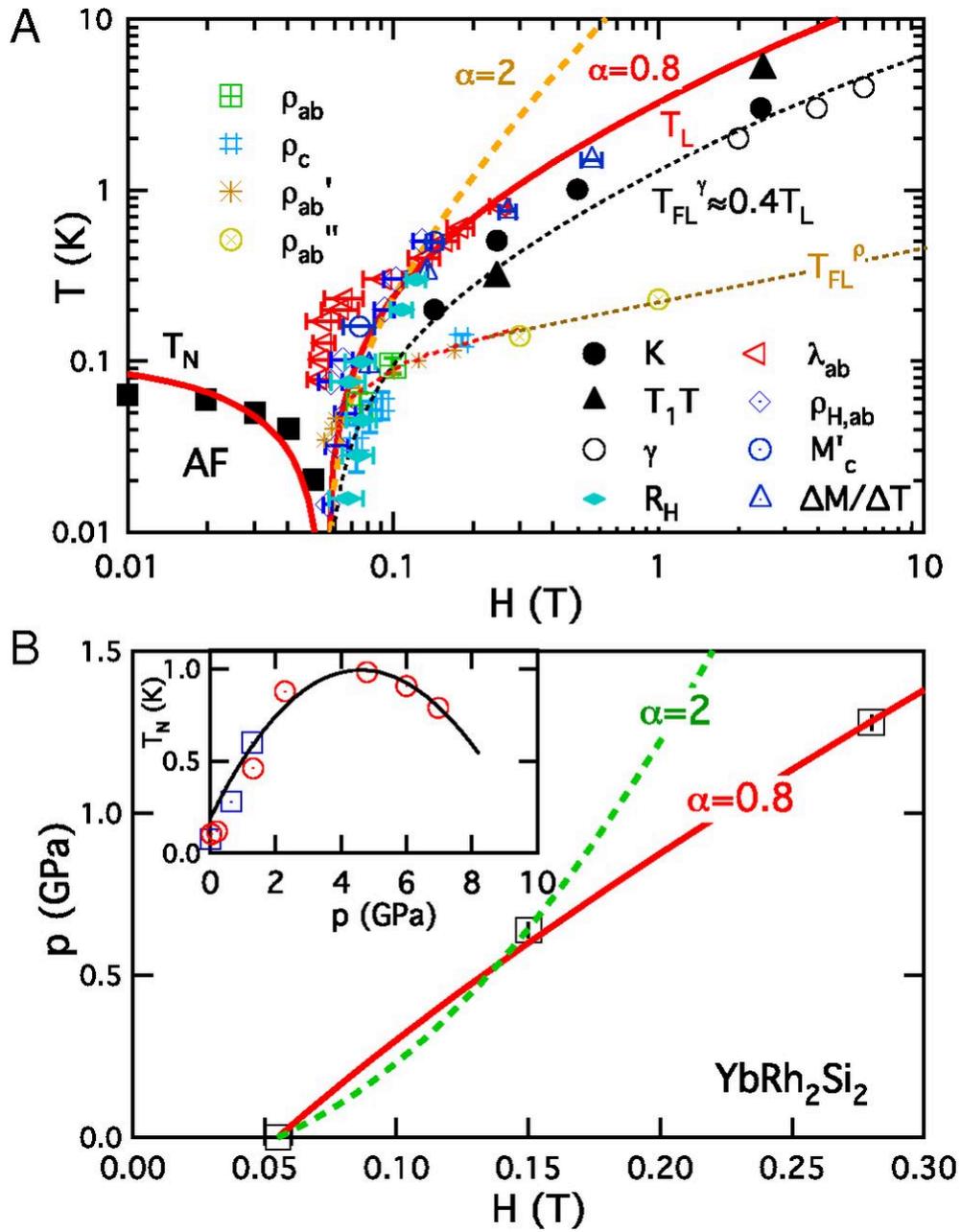

Fig. 10. Phase diagram of YbRh$_2$Si$_2$. (A) and (B) as in Fig. 9, but for an antiferromagnetic rather than superconducting quantum phase transition.[35]

## IV. Toward a New Microscopic Theory

### 1. Collective hybridization in the Kondo lattice model

In this section we develop more fully the ideas presented in Sec. II and, in particular, suggest a mechanism for collective hybridization missing in current discussions on the Kondo lattice. Recall first the conventional description of the Kondo lattice in terms of the Doniach phase diagram shown in Fig. 11. This is characterized by the Kondo temperature $T_K \sim \rho^{-1}\exp(-1/\rho J)$ and the RKKY or Heisenberg interaction $J_H \sim \rho J^2$, in the notation of Sec. II.

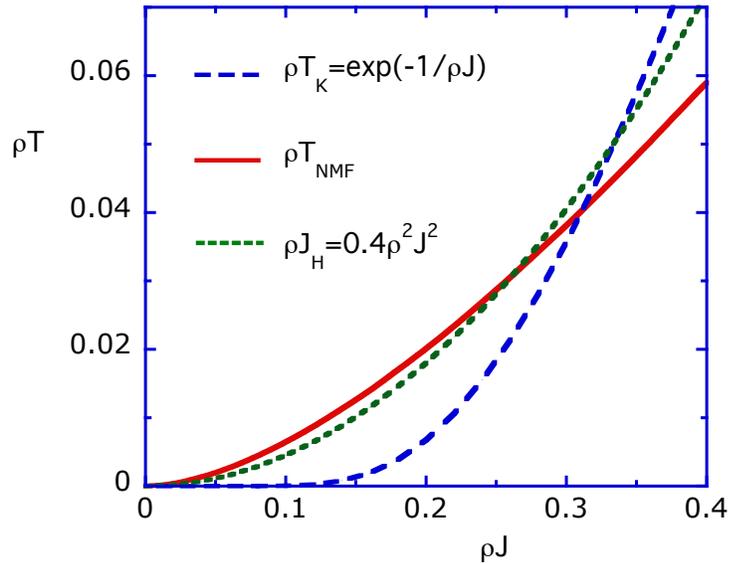

Fig. 11. The dependence on the Kondo interaction parameter, $J$, of the Kondo temperature, $T_K$, and the Néel temperature, $T_{NMF}$, in a slave boson based mean field theory of the Kondo Hamiltonian.[37] $T_K$ is obtained assuming that the antiferromagnetic order parameter is zero and $T_{NMF}$ assuming the hybridization order parameter is zero. $T_K$ is the same as that in the single-ion problem. The Heisenberg interaction parameter $J_H$ shown is a $(\rho J)^2$ fit to $T_{NMF}$. We note that the dependence of $J_H$ on $J$ is similar to that observed for the measured coherence temperature $T^*$ (see Eq. 2). When both order parameters are allowed to vary to minimize the total free energy, the Néel temperature follows $T_{NMF}$ up to $T_{NMF} \sim T_K$ and then collapses to zero. A finite value of both order parameters exists over a narrow region in $\rho J$ (i.e., in the range $0.3 \le \rho J \le 0.4$).

Magnetic order is assumed to develop below a temperature of the order of $J_H$ in the regime where $T_K$ is below $J_H$, and a magnetic quantum critical point is expected to arise at a critical coupling, $J_c$, where $T_K \sim J_H$. This phase diagram thus consists of a magnetic transition line $T_N > T_K$ for $J < J_c$, and a crossover line $T_K > J_H$ for $J > J_c$. The later regime is characterized by a disordered local moment state above $T_K$ and a coherent hybridized conduction electron and f-electron state below $T_K$.

This picture can be made concrete in the slave boson mean field approximation in which the hybridized state is represented in terms of a lower and upper band as shown in Fig. 12. [4] This is for the symmetric Kondo lattice model of a single conduction band coupled to a doubly degenerate f level in which double occupancy is forbidden. (The f electrons are then described as pseudo fermions or spinons.) The indirect gap is twice $T_K$, which in this model is the same as the single-ion Kondo temperature, while the direct gap, $\Delta = 2\sqrt{T_K/\rho}$, is twice the geometrical average of $T_K$ and the conduction electron half-bandwidth $1/\rho$; thus the direct gap can be much greater than the indirect gap.

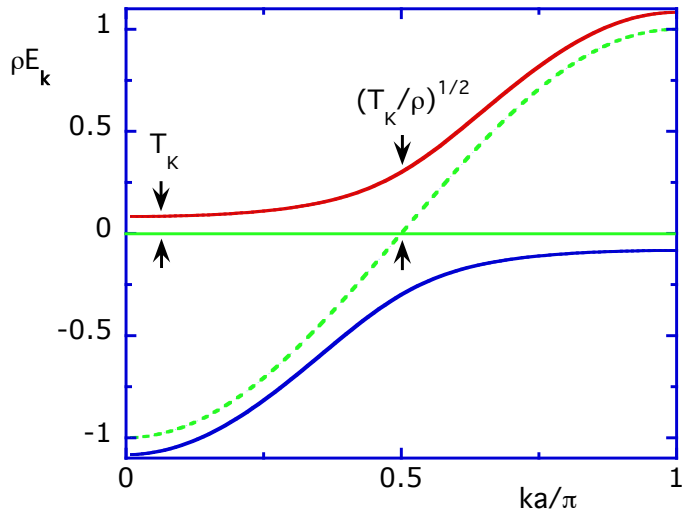

Fig. 12. Hybridized conduction and f-electron bands in the slave boson mean field theory for the Kondo lattice in the symmetric case where the centers of both bands coincide.[4] In this model $T_K$ is the same as the Kondo temperature for a single-ion. The indirect gap is $2T_K$ and the direct cap is $2\sqrt{T_K/\rho}$. Both gaps are predicted to vanish at a temperature of order $T_K$. Fluctuations around the mean field description lead to pronounced broadening of the bands (except near to Fermi level in the ground state) and to a gradual rather than sharp suppression of hybridization with increasing temperature.

In this mean field model $\Delta$ is a temperature dependent order parameter that vanishes at a critical temperature of the order of $T_K$. Thus, the hybridized state below $T_K$ and the unhybridized state above $T_K$ are separated by a well-defined critical point. Although a sharp phase transition is not expected to arise when fluctuations around the mean field state are included, the scale $T_K$ is still thought to be relevant as a measure of the crossover temperature between the nominally hybridized and nominally unhybridized state, below and above $T_K$, respectively.

As we have seen in Sec. III, analyses using the two-fluid model have uncovered a breakdown of the predictions of the above theory and of the Doniach phase diagram in general. In particular the coherence temperature $T^*$, at which hybridization begins to develop on cooling from the high temperature incoherent state, is found to be much larger than $T_K$ as defined above, and near to magnetic quantum critical points investigated scales as $J_H$, i.e.,

$$T^* \sim J_H \gg T_K \qquad [15]$$

Moreover, it is found that the critical value of the dimensionless Kondo coupling, $\rho J_C$, is typically a factor of two lower than that predicted by slave boson based mean field models that include both the hybridization and antiferromagnetic order parameters self consistently.[1][37]

These findings indicate that the conventional slave boson mean field description, although instructive in several respects, may fail to capture an important difference between single-ion hybridization and lattice hybridization, a difference that is expected to disappear only in the limit of high f-level degeneracy. This limit may be misleading for the systems of interest here that have appreciable crystal field splittings and hence low effective degeneracies.

To check on the possible role of intersite contributions to hybridization, we consider a different type of mean field theory that is not restricted to high f-level degeneracy. This is the Gutzwiller mean field approximation, which indeed predicts a strong lattice enhancement of the hybridization.[38][39] In particular for an f-level degeneracy of two, it leads to the same hybridized band picture as that shown in Fig. 12, but with $T_K$ replaced by

$$T_K^* = \rho^{-1}\exp\left(-\frac{1}{2\rho J}\right) = \sqrt{T_K/\rho} \gg T_K \qquad [16]$$

Thus, the indirect gap is $2T_K^*$ and the direct gap is $2\sqrt{T_K^*/\rho}$, both much larger than that predicted by the slave boson mean field approximation.

The enhanced scale $T_K^*$ tends to reduce the critical Kondo coupling at the magnetic quantum critical point and in the vicinity of the latter we expect $T_K^*$ to be of the order $J_H$ (Fig. 13). Since $T_K^*$ marks the onset of hybridization, it may be associated with the coherence temperature $T^*$. Thus, the Gutzwiller mean field theory would seem to lead to

$$T_K^* \sim T^* \sim J_H \gg T_K \qquad [17]$$

and a direct hybridization gap of the order $\sqrt{T^*/\rho}$, rather than $\sqrt{T_K/\rho}$. These predictions, namely (i) a reduced critical coupling, $\rho J_c$, (ii) an enhanced coherence temperature, $T^*$ tracking $J_H$, and (iii) an enhanced hybridization gap, appear to be in order of magnitude agreement with observation.

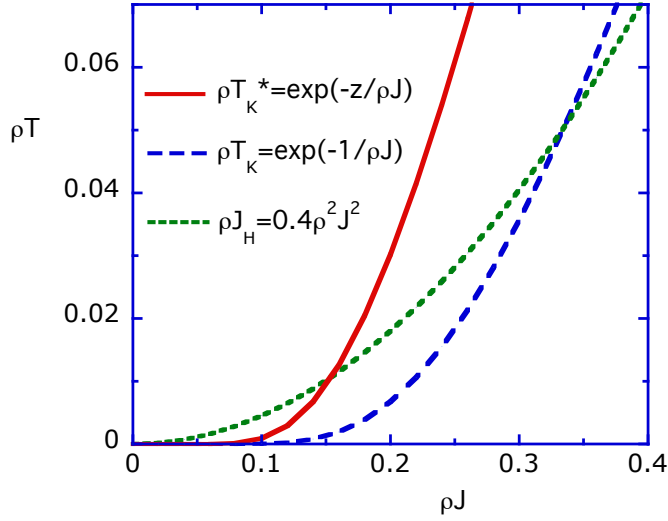

Fig. 13. Similar to Fig. 11, but including the $\rho J$ dependence of the enhanced temperature scale $T_K^*$ arising from collective hybridization in the Gutzwiller[38][39] as compared with the slave boson mean field models. The parameter $z$ tends to unity in the limit of large f-level degeneracy, $N$, and to ½ in the limits of large spatial dimensions for $N = 2$; more generally, $0.5 < z < 1$. The $T_K^*$ line in the plot is for an intermediate value, $z = 3/4$. This leads to a suppression of the magnetic quantum critical point by about a factor of two, yielding a value of $\rho J_c$ in agreement with observation. Near to the quantum critical point $T_K^* \sim J_H \gg T_K$, also as observed in many Kondo lattice systems (Table I).

The Gutzwiller mean field theory is expected to overestimate the lattice enhancement in systems of low spatial dimensions. However, numerical analyses suggest that the enhancement remains appreciable even in dimensions as low as two. [40][41][42]

An examination of the Gutzwiller approximation shows in a formal way how the lattice enhancement arises. [38][39] The effect can be traced back, in the symmetric case, to the form of the trial many-particle wavefunction before projection, i.e., before removal of configurations in which an f level is doubly occupied. The difference between hybridization for the lattice and a single ion shows up in the use of Block f states in the former case and single site local f states in the latter case. The inclusion of Block f states in this way leads to intersite correlations and a lattice enhancement of the hybridization.

Although this model appears to shed light on how lattice enhancement can arise, it fails to treat hybridization and magnetic correlations on equal footing and therefore does not provide an appropriate microscopic description of the two-fluid model. An alternative approach might be to consider, as discussed in Sec. II, a Hamiltonian, which includes not only the usual on-site Kondo coupling but also an intersite magnetic (Heisenberg) interaction. To permit the use of a simple mean field analysis it would also seem to be necessary to introduce an additional coupling to represent in a transparent way the idea of lattice enhanced or collective hybridization.

For a realistic description it is also necessary to take account of the effect of the existence of a multiplicity of f levels. Even in the presence of a crystal field splitting several times greater than $T^*$ the effective value of $J$ that determines $T_K$ can be significantly greater than that determining the RKKY coupling, or $J_H$,[4] which has the effect of reducing the dimensionless coupling, $\rho J_c$. Other factors that can reduce $\rho J_c$ include complexities of the conduction electron spectrum and lattice frustration. Numerical analyses thus far seem to exclude these effects and therefore predict unrealistically high values of $\rho J_c$.[40]

In the next section we review the emergence of enhanced or collective hybridization within the two-fluid model and consider yet another possible way forward for understanding, in particular, the regime in which both hybridized and unhybridized states in some sense coexist, *i.e.*, in the Kondo liquid two-fluid regime.

**2. Collective hybridization and the two-fluid model**

It is instructive to discuss collective hybridization in more phenomenological terms, beginning with the simplest mean field description of the coupled conduction electron and f-electron spins. If $\chi_{ff}^0(q)$ and $\chi_{cc}^0(q)$ are the non-interacting wavevector dependent magnetic susceptibilities of the f electrons and conduction electrons, respectively, then in the presence of a Kondo coupling $J$, which yields the RKKY Heisenberg interaction $J_H = J^2 \chi_{cc}^0(q)$, the interacting susceptibilities in mean field theory without hybridization are

$$\chi_{ff}^{MF}(q) = \frac{\chi_{ff}^0(q)}{1 - J_H \chi_{ff}^0(q)} \qquad [18]$$

$$\chi_{cc}^{MF}(q) = \frac{\chi_{cc}^0(q)}{1 - J_H \chi_{ff}^0(q)} \qquad [19]$$

On substituting $\chi_{ff}^0(q) = \chi_{ff}^0 = C/T$ for a local moment, where $C$ is the Curie constant, we them have a Kondo interaction driven instability of the coupled spin system at a wavevector $q^*$ where $J_H(q)$ is maximum and at a temperature $T_{NMF} = C J_H(q^*)$. (Note that $\chi_{ff}^0$ is the same as $\chi_0$ in Sec. III.5.)

This analysis is, however, incomplete in particular because it does not include hybridization of conduction electron and f-electron states that is also brought about by the Kondo coupling $J$. Within the two-fluid model the effect of hybridization is to reduce the strength of the local-moment contribution by the factor $(1 - f(T))$, so that

$$\chi_{ff}(q) = \frac{(1-f(T))\chi_{ff}^0(q)}{1 - J_H(1-f(T))\chi_{ff}^0(q)} \qquad [20]$$

In this model the Néel temperature, $T_N$, is defined by the singularity of $\chi_{ff}(q^*)$, and is hence given by $T_N = C J_H(q^*)(1-f(T_N))$ for $f_0 < 1$. Thus the inclusion of hybridization suppresses Néel order, $T_N < T_{NMF}$, at least if $T^* > T_{NMF}$ for the assumed form of $f(T)$ (Eq. [5]). Also, the delocalization temperature, $T_L$, is defined by the singularity of the inverse of $\chi_{ff}(q^*)$, which yields Eq. [7] for $f_0 > 1$.

This is a description within the two-fluid model of the dual role of the Kondo coupling $J$ in promoting both magnetic polarization and hybridization. As we have emphasized, however, the conventional description (specifically, the slave boson mean field theory) underestimates the role of hybridization in the lattice, a deficiency that we have dealt with in our analyses in terms of phenomenological parameters.

We have emphasized the importance of including enhanced collective hybridization in any microscopic model of Kondo lattice behavior. In the last section we have indicated some possible ways forward. Here we add one more consideration that may prove fruitful, namely that collective hybridization is not a uniform phenomenon in the lattice, but arises in wave-like patterns analogous to density waves in more conventional forms of order or to still more exotic textured and dynamical states as suggested in Sec. II.

**V. What We Have Learned**

Before concluding we pause to summarize what we have learned from our examination of experiment, phenomenology, and microscopic theory.

* On the Kondo lattice, there are significant differences in how a lattice of local moments and a single-ion local moment hybridize with conduction electrons.

* The collective hybridization for a lattice of local moments is generally much more effective than single-ion local hybridization: it begins at a crossover temperature.

$$T^* \sim J_H \sim \rho J^2 \gg T_K \sim \rho^{-1}\exp(-1/\rho J) \qquad [21]$$

instead of $T_K$.

* The direct hybridization gap, $\Delta$, reflects the collective origin of hybridization; experiment shows that $\Delta \sim (T^*/\rho)^{1/2}$ instead of $\Delta \sim (T_K/\rho)^{1/2}$ predicted by slave boson mean field theory.

* The quantum critical point of heavy electron materials is not set by the competition between the Kondo (single-ion) and RKKY energy scales given in the Doniach phase diagram; it is found in the vicinity of $\rho J_c \sim 0.17$, significantly below the Doniach value of $\rho J_c \sim 0.35$.

* The two-fluid description makes it possible to use hybridization effectiveness as a quantum tuning parameter and disentangle the average behavior of the itinerant heavy fermion liquid and local moment spin liquid below $T^*$.

* It yields a simple expression for the relative strength, $f(T)$, of the emergent itinerant heavy fermion liquid in terms of a hybridization effectiveness parameter, $f_0$, that is defined to be unity at the quantum point, $f(T) = f_0 (1 - T/T^*)^{3/2}$.

* When $f_0 > 1$, hybridization ends at finite temperatures along a delocalization line, $T_L$, given by $f(T_L) = 1$ or $T_L = T^* \left(1 - f_0^{-2/3}\right)$.

* The itinerant heavy electron liquid that emerges below $T^*$ is a new state of matter, the Kondo liquid, a non-Landau Fermi liquid that can be seen directly in measurements of the Knight shift that yield its emergence and subsequent temperature variation.

* The behavior of the Kondo liquid is universal with a quasiparticle density of states, $N_F(p,T^*)$, that is logarithmically divergent as it approaches the quantum critical point; in the two-fluid description it depends only on the hybridization effectiveness parameter, $f_0$, and $T^*$, with

$$N_F(p,T) = \frac{3\ln 2}{2\pi^2 T^*(p)} f_0(p)\left(1 - \frac{T}{T^*}\right)^{3/2}\left(1 + \ln\frac{T^*}{T}\right)$$ [22]

* The circumstances under which superconductivity emerges from the Kondo liquid and the superconducting transition temperature, $T_c$, are determined by the quantum critical tuning parameter, $f_0$, while a phenomenological BCS-like theory appears capable of explaining the variation of $T_c$ with pressure for both CeCoIn$_5$ and CeRhIn$_5$.

The relative strength of the local moment spin liquid that coexists with the Kondo liquid is $(1 - f(T))$: hybridization reduces its Néel temperature, $T_N$, dramatically, from $J_{RKKY} \sim T^*$ to $\sim (1 - f(T_N))T^*$, while the strength of its magnetic moments is reduced by a factor of $\sqrt{1 - f(T)}$.

* There are several mechanisms that could lead to lattice enhancement of hybridization of the conduction electrons and f electrons, *i.e.*, to collective hybridization in our terminology, and promising candidates have been proposed here.

**VI. Concluding Remarks**

We have presented in this paper an experiment based manifesto in which we have identified terms missing from earlier attempts to develop a microscopic theory and proposed several scenarios for including these in future theoretical approaches. Our suggested approach to a microscopic treatment of collective hybridization is based on the experiments and the two-fluid phenomenology that are described in Sec. III; these provide us with the basic concepts whose inclusion in future microscopic models should lead to results that both explain existing experiment and suggest new ones.

In considering the current status of theory and experiment on heavy electron materials, it is instructive to look back at the history of superconductivity. By December 1956, one had in place the following:

* The key physical ideas to which Fritz London had made seminal contributions- the two-fluid model and the primacy of the Meissner effect;

* Bardeen's phenomenological description based on the emergence of a quasi-particle energy gap in the superconducting state;

* The Bardeen-Pines candidate for the origin of superconductivity based on the isotope effect - a screened attractive interaction between electrons near the Fermi surface that comes about when the attractive phonon-induced (Frohlich) electron interaction wins out over their screened Coulomb repulsion;

* Cooper's toy model for how that interaction led to an instability of the Fermi surface.

The stage was set for BCS.

It is our hope that developments described in this paper have brought us one step closer to a similar stage with heavy electron materials and that a microscopic theory of heavy electron emergence and Kondo liquid behavior will soon be forthcoming.

**Acknowledgements**

This work was performed in part at the Aspen Center for Physics, which is supported by National Science Foundation (NSF) grant no. PHY-1066293. We thank the Centre, NSF and many colleagues at the Centre and elsewhere for stimulating discussions. G.L. also acknowledges support from the Engineering and Physical Sciences Research Council (EPSRC) grant no. EP/K012894/1, and from the Cavendish Laboratory and Trinity College, Cambridge. Y.Y. was supported by the State Key Development Program for Basic Research of China (grant no. 2015CB921303) and the Strategic Priority Research Program of the Chinese Academy of Sciences (grant no. XDB07020200).

**Appendix I. Some Suggested Experiments**

Future experiments can clarify the basic concepts presented here, give additional information on the parameters that enter the two-fluid phenomenology, and offer further guidance in developing a microscopic theory. The following appear to us to be of particular interest.

*The delocalization line.* A direct measure of the crossover at $T_L$ is highly desirable, since to date we have had to rely on indirect determinations of its existence. The most direct determination would involve seeing in a magnetic or other measurement the crossover from two component to one component behavior that must take place because at or below $T_L$ there is no trace of local moment behavior. Because it has proved difficult to date to measure the bulk susceptibility at finite pressures, it would seem easiest to measure the field dependence of $T_L$ by carrying out simultaneous measurements of the Knight shift and the uniform bulk magnetic susceptibility over a range of magnetic fields around the magnetic quantum critical point. A good place to start might be with simultaneous Knight shift and bulk susceptibility measurements on Ce115 at fields in the vicinity of 4 Tesla, to verify the conjecture that the maximum in resistivity seen by Zaum *et al.*[11] is a measure of $T_L$.

*Magnetic field induced changes in Kondo liquid behavior.* Jiao *et al.*[43] have recently studied the phase changes in Rh115 induced by the application of strong external magnetic fields. They find that at ambient pressure local moment behavior disappears and one has a jump in the size of the Fermi surface along a line in the temperature magnetic field plane that ends at a localization quantum critical point at $B \sim 30T$, all within the AF ordered state. To the right of this line where one has a Kondo liquid, it exhibits itinerant antiferromagnetic order over a considerable range of fields and temperature. It will be interesting to study the behavior of both $T_L$ and the Kondo liquid at temperatures greater than $T_{Néel}$.

*The Quantum Critical Fermi Liquid.* Experiment has shown that at ambient pressure and zero external field CeCoIn$_5$ is not far from a quantum critical point, since application of a pressure of ~ 1.1 GPa or a magnetic field of ~ 4.1 Tesla will bring the material to the QCP. As noted earlier, Zaum *et al.*[11] have been able to identify two distinct regions of quantum critical behavior in the vicinity of its magnetic quantum critical point, but we do not yet know from experiment whether these exist in the vicinity of the quantum critical pressure. A closely related question is mapping out a possible crossover from Kondo liquid to quantum critical behavior for the Fermi liquid found in weakly hybridized materials that is suggested by both specific heat and quantum oscillation (dHvA) experiments.

*The fate of the remnant heavy electrons and the role played by lightly hybridized electrons.* A number of experiments have suggested that within the antiferromagnetic state there continue to be a number of remnant heavy electrons. In CeRhIn$_5$, this is shown by the specific heat coefficient of ~ 56 $\mathrm{mJ/mol\ K^2}$ that is much larger than that of pure conduction electrons. It is then natural to wonder about the fate of these remnant heavy electrons: whether they will condense into superconductivity or eventually become a Fermi liquid. On the other hand, for materials with multiple Fermi sheets, there may also exist lightly hybridized electrons at low temperatures. It is desirable to see if these different types of charge carriers can be separated and their role identified in experiment.

*Quadrupolar Quantum Criticality*. The study by quantum oscillatory effects in high magnetic fields of the evolution from the hybridized to the unhybridized state is in practice hampered by the tendency of the applied magnetic field to suppress the Kondo liquid state in favor of a more conventional Fermi liquid state. This is less likely to arise in systems in which the f-electron doublet involved in the Kondo effect is non-magnetic in nature, as for example in the case of a quadrupolar doublet near a quadrupolar QCP. An example is UBe$_{13}$, which is believed to be on the border of a quadrupolar QCP and exhibits electronic and magnetic properties that are very weakly dependent on magnetic field. Other examples of quadrupolar quantum critical points are being discovered and provide new opportunities for investigating the evolution of the Fermi surface with temperature and more generally for testing the applicability of the concept of elementary excitations near to a quantum phase transition.

*Transitions from large to small Fermi surfaces as a function of temperature.* Systems with quadrupolar f-doublets are ideally suited for the above studies, but involve new subtleties and complexities compared with the simpler heavy fermion system of main interest in this review that may be described in terms of f-electron magnetic doublets, or effectively spin doublets. In these systems the crossover from the hybridized to the unhybridized state (above $T_L$) may be investigated by controlling both the applied pressure and magnetic field so that the border of antiferromagnetism may be tuned to be in the vicinity of the experimental magnetic field and temperature range where quantum oscillations can be detected. In this regime the crossover temperature $T_L$ can be tuned to be almost arbitrarily small and, in particular, sufficiently small so that quantum oscillations can be detected well above $T_L$. This type of study would seem to be possible not only in the AF state, but also in the non-AF state of, for example, $CeRhIn_5$.[43] Such an investigation may be aided by the fact that the mass of the quasiparticles is expected to be reduced in the unhybridized state above $T_L$. This could lead to a dramatic temperature dependence of the quantum oscillation amplitude where "large" sheets of the Fermi surface with high mass, and hence strongly decreasing quantum oscillation amplitude with increasing temperature, would give way to "small" sheets of the Fermi surface with small mass at high temperatures, with weakly temperature dependent quantum oscillation amplitude that might be observed far above $T_L$.

Scattering of conduction electrons from thermal fluctuations would naively be expected to strongly attenuate the quantum oscillation amplitude above realistic values of $T_L$. However, it has been shown that in contrast to scattering from quenched disorder the scattering from thermal fluctuations can in some cases have little effect on the quantum oscillation amplitude. This has been proved experimentally and theoretically for the case of the electron-phonon interaction. A similar unexpectedly weak effect of thermal fluctuations might apply to the cases of interest here, making the proposed experiment of particular interest.

*Scanning tunneling microscopy* has proven to be a powerful tool for studying inhomogenities in strongly correlated electron materials so it is natural to ask whether it will prove capable of detecting the fluctuating regions of hybridization that we have suggested might exist between $T^*$ and $T_L$, and whose average values are plausible candidates for the fluids that appear in the two-fluid phenomenological model. We note that we have not considered the existing STM experiments on the emergence of the Kondo liquid, such as that of Aynajian *et al.*,[44] on $CeCoIn_5$ material, because at present the interpretation of STM results is model dependent, and to date their results have not been interpreted using a collective model that incorporates hybridization waves at some finite wavevector.

**Appendix II. The Cuprate Connection**

What we have learned about heavy electron emergence may also be applicable to the coupling between d-electron localized spins and p-band holes in the cuprate superconductors. Consider, for simplicity, the region of hole doping that lies beyond the AF domain, and focus on the system behavior in the Cu-O plane. There, at high temperatures, one will have a d-electron local moment spin liquid that is Kondo coupled to a background of p-band holes, a physical situation identical to that we have been considering for Kondo lattice materials.[1] As the temperature is lowered these will begin to hybridize, most likely at a temperature set by the effective doping-dependent interaction between the local moments that can be determined using fits to the susceptibility of the 2d Heisenberg model[45]. The result will be two fluids whose properties have been analysed in some detail by Barzykin and Pines [hereafter BP][46][47] using a two-fluid description borrowed from that developed for heavy electrons.

---

[1] In contrast to typical Kondo lattice systems, in this case the conduction band is nearly full so that the Fermi surface of the lower hybridized band is close to half filling. This had made it possible in some cases to describe the problem in terms of a single nearly half-filled band picture.

The BP analysis suggests that the two fluids coexist at doping levels that lie well beyond optimal doping, while their coexistence at optimal doping has been verified in the seminal experiments of Slichter *et al.* and Haase *et al.*[48][49]. The BP analysis also shows that the local moment spin liquid displays a simple doping dependent behavior, characterized by a temperature-independent order parameter, a result that suggests that the Kondo hybridization has ended at a temperature that is somewhat greater than that over which most measurements have been carried out, 300K and below. That, together with the fact that the itinerant fermion effective mass enhancement would be small, since it varies as $1/T^*$ and $T^*$ is large, would explain why there is no evidence thus far that the itinerant fermion liquid displays Kondo liquid behavior.

What the experiments of Slichter *et al.* tell us is that the pseudogap behavior should be regarded as an instability of the Kondo-like system, that it is only in the itinerant fermion (or hole) liquid that superconductivity emerges, and that a microscopic theory based on the one-band Hubbard model will not fully capture the two-fluid physics that appears likely to play an important role in determining the effective interaction between itinerant fermions (or holes) that gives rise to their superconductivity.